\documentclass{article}
\usepackage{arxiv}

\usepackage[utf8]{inputenc} % allow utf-8 input
\usepackage[T1]{fontenc}    % use 8-bit T1 fonts
\usepackage[numbers]{natbib}
\usepackage{hyperref}       % hyperlinks
\usepackage{url}            % simple URL typesetting
\usepackage{booktabs}       % professional-quality tables
\usepackage{amsfonts}       % blackboard math symbols
\usepackage{nicefrac}       % compact symbols for 1/2, etc.
\usepackage{microtype}      % microtypography
\usepackage{lipsum}		% Can be removed after putting your text content
\usepackage{graphicx}
\usepackage{doi}
\usepackage{amsmath}
\usepackage{eucal}

\usepackage{siunitx}
\newcommand{\LP}[2][]{$\mathrm{LP}_{#2}^{#1}$}

\title{Soliton trapping and orthogonal Raman scattering in a~birefringent microstructured fiber}

\author{
    {Karolina Stefańska}\\
	Department of Optics and Photonics\\
	Wrocław University of Science and Technology\\
	\texttt{karolina.stefanska@pwr.edu.pl} \\
	\And
	{Sylwia Majchrowska} \\
	Department of Optics and Photonics\\
	Wrocław University of Science and Technology \\
	\texttt{sylwia.majchrowska@pwr.edu.pl} \\
	\And
	{Karolina Gemza}\\
	Department of Optics and Photonics\\
	Wrocław University of Science and Technology\\
    \texttt{karolina.gemza@student.pwr.edu.pl} \\
	\And
	{Grzegorz Soboń} \\
	Laser \& Fiber Electronics Group, \\
    Wrocław University of Science and Technology, \\
	\texttt{grzegorz.sobon@pwr.edu.pl} \\
    \And
	{Jarosław Sotor} \\
	Laser \& Fiber Electronics Group, \\
	Wrocław University of Science and Technology\\
	\texttt{jaroslaw.sotor@pwr.edu.pl} \\
	\And
	{Paweł Mergo} \\
	Laboratory of Optical Fiber Technology, \\
    Maria Curie-Skłodowska University, \\
    \texttt{pawel.mergo@mail.umcs.pl} \\
	\And
	{Karol Tarnowski} \\
	Department of Optics and Photonics\\
	Wrocław University of Science and Technology\\
	\texttt{karol.tarnowski@pwr.edu.pl} \\
	\And
	{Tadeusz Martynkien}\\
	Department of Optics and Photonics\\
	Wrocław University of Science and Technology\\
	\texttt{tadeusz.martynkien@pwr.edu.pl} \\
}

\begin{document}
\maketitle

\begin{abstract}
We report on trapped pulse generation in birefringent microstructured optical fiber.
Linearly polarized fs pulses are injected into the microstructured fiber in anomalous dispersion regime.
We observed experimentally that soliton pulse polarized along the fast fiber axis partially transfers its energy to the orthogonal polarization.
The generated pulse is amplified through the orthogonal Raman gain.
The two polarization components are located at group-velocity matched wavelengths.
The experimental works are supported with numerical simulations.
The obtained results are important for the light sources using self-frequency shifted solitons in applications demanding high polarization purity.
\end{abstract}

% keywords can be removed

\noindent The optical solitons are intense light pulses that propagate in a nonlinear medium in anomalous dispersion regime. They retain shape due to balance between chromatic dispersion and self-phase modulation.
The possibility of transmission of such undistorted pulses in optical fibers was predicted by Hasegawa and Tappert in 1973~\cite{Hasegawa1973}.
In 1986, Mitschke and Mollenauer discovered that an ultrashort soliton undergoes self-frequency shift caused by energy transfer from the higher to the lower frequency part of the spectrum due to Raman scattering \cite{Mitschke1986} (SSFS -- soliton self-frequency shift).
This phenomenon allows for continuous tuning of the soliton's spectral position and facilitates reaching wavelengths not available with typical laser sources.
In recent years, the possibility of effectively tuning solitons in photonic crystal fibers was investigated.
The photonic crystal fibers appeared to be perfect media for spectral tuning of solitons as they provide control over the shape of dispersion curve and allow to obtain a small effective mode area which results in high effective nonlinearity~\cite{Takayanagi2008,Pant2010,Sobon2017,Sobon2018}.

The second thread of the recent research related to solitons in optical fibers is focused on few-mode fibers (FMFs) and multimode fibers (MMFs)~\cite{Picozzi2015}.
In reference to photonic crystal fibers, the FWFs and MMFs have relatively high effective area, thus they can guide high-intensity pulses.
Moreover, they enable more complex dynamics by providing additional degrees of freedom to the system.
The interest in multimode fibers is also motivated by the possibility of increasing the capacity of optical telecommunication networks by utilizing spatial division multiplexing.
Propagation of solitons in higher-order modes of MMF was investigated by Rishøj et al.~\cite{Rishoj2019}.
They demonstrated an intermodal soliton conversion in a step-index multimode fiber initiated by Raman scattering.
They observed series of self-soliton conversions resulting in transfer from initially excited \LP{0,19} mode at \SI{1045}{\nano\meter} to \LP{0,15} mode at \SI{1587}{\nano\meter} on \SI{12}{\meter} of propagation.
The conversion was enabled by Raman scattering and group velocity matching of distinct spatial modes.

The Raman scattering is not the only phenomenon that is important to describe nonlinear pulse propagation.
The next mechanism is the cross-phase modulation (XPM).
Due to XPM, the intense pulse can force the common group velocity causing trapping of the other copropagating pulse.
When two copropagating pulses are orthogonally polarized, the XPM induces nonlinear phase shifts in both polarization components which depend on the intensity of the orthogonal polarization component \cite{Agrawalbook}.
This kind of mutual interaction can lead to the trapping of orthogonally polarized pulses.
The first studies on soliton trapping showed a numerical investigation of XPM in conventional birefringent fibers \cite{Menyuk1987, Menyuk1988}.
They were followed by the experimental demonstration of soliton trapping in low-birefringent optical fiber in the low power regime \cite{Islam1989}.
In this experiment, the linearly polarized pulse with azimuth angle $\theta = \SI{45}{\degree}$ excited two polarization components.
The slow-axis component lowered its wavelength to increase the group velocity, and the fast-axis component raised its wavelength to decrease the group velocity.
Finally, both components propagated with a~common group velocity.

Interestingly, there exists a mechanism that can transfer energy from one polarization component to the other.
This mechanism is provided by orthogonal Raman scattering.
The energy from the shorter wavelengths in one polarization is transferred to the longer wavelengths in orthogonal polarization.
Although the orthogonal Raman response is weaker than co-polarized Raman response~\cite{LinAgrawal2006}, it plays a key role in the pulse trapping observed in the conventional birefringent fiber by Nishizawa \cite{Nishizawa2001, Nishizawa2002a}.
In those experiments, the soliton polarized along the slow fiber axis transferred part of its energy to the fast axis.
The generated pulse polarized along the fast fiber axis was trapped by slow-axis component and feeded by the trapping pulse through the Raman scattering.
The authors confirmed that the pulses overlapped temporally and co-propagated along the fiber.
In the following work, the authors also investigated pulse trapping of co-linearly polarized pulses across zero-dispersion wavelength \cite{Nishizawa2002b}.
Additionally, pulse trapping of an orthogonally polarized continuous wave \cite{Shiraki2010} and even pulse trapping and amplification of incoherent light from a super-luminescent diode \cite{Shiraki2012} were demonstrated in birefringent fibers.

In the referenced works~\cite{Nishizawa2001, Nishizawa2002a, Shiraki2010, Shiraki2012}, the conventional birefringent fiber was used.
In this type of fiber, the group velocity matching of polarization modes is possible between a shorter wavelength aligned to the slow axis and a longer wavelength aligned to the fast axis.
Consequently, the pulse trapping is possible when trapping pulse is aligned to the slow axis and the generated trapped pulse is aligned to the fast axis.
Simultaneously, pumping in the fast axis is preferable in conventional birefringent fibers to avoid gradual pulse depolarization by orthogonal Raman scattering.

In microstructured birefringent fibers, the situation is opposite.
The distinguishing feature of microstructured fibers is negative value of group birefringence which means that the pulses polarized along the fast fiber axis have lower group velocity than pulses polarized along the slow fiber axis.
Consequently, a shorter wavelength aligned to the fast axis and a longer wavelength aligned to the slow axis can be group-velocity-matched.
Our aim was to investigate the influence of orthogonal Raman scattering on solitons propagating in birefringent microstructured optical fiber. 
To our knowledge, this is the first detailed study of soliton trapping in optical fiber with negative group birefringence.

To observe the effect of Raman scattering on solitons, we used mode-locked Er-doped fiber laser (FemtoFiber pro IR, TOPTICA Photonics) operating at \SI{1.56}{\micro\meter} and generating \SI{23}{\femto\second} pulses with a repetition rate of 80 MHz and average output power of about \SI{200}{\milli\watt}.
To control the~power level and the~polarization state of the laser pulses, they were passed through a polarizer and half-wave plate.
For measuring an average beam power, we used a~thermal power sensor (Thorlabs, S401C).
The pulses were then injected into the fiber with an aspheric lens (Thorlabs, \makebox{C230TMD-C}).

\begin{figure}
    \centering
    \includegraphics[width=.7\linewidth]{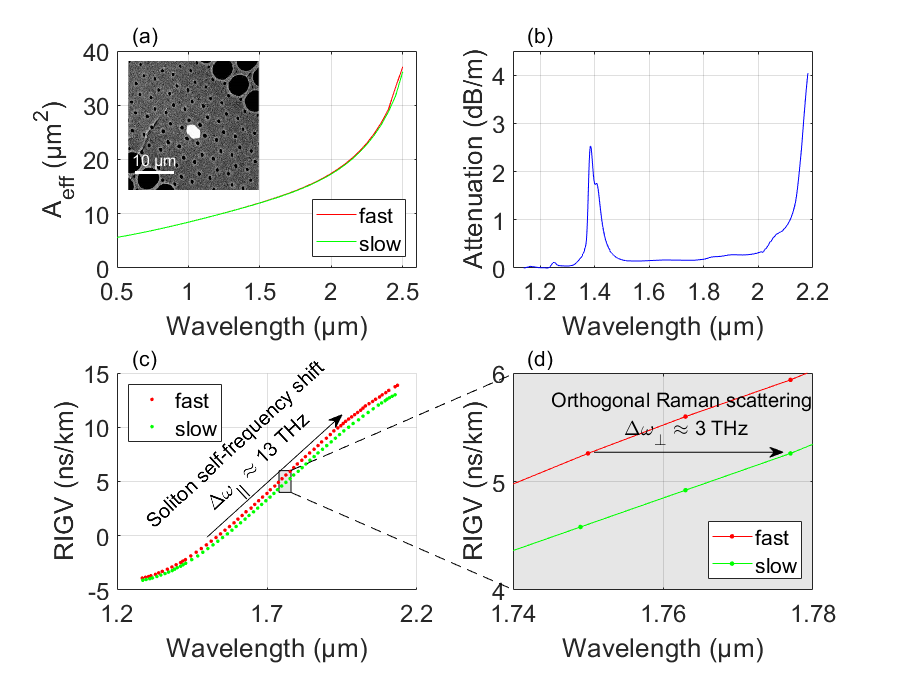}
    \caption{Parameters of the microstructured fiber. Calculated mode effective area (a), measured attenuation (b) and relative inverse group velocity of the two polarization modes (c,d) with indications of the characteristic frequency shifts responsible for the given mechanism. Inset in (a) shows the SEM image of fiber's cross-section.}
    \label{fig:fiber}
\end{figure}

\newpage
We used an in-house developed microstructured fiber fabricated using stack-and-draw method.
The SEM image of the fiber's cross-section is shown in the inset of Fig.~\ref{fig:fiber}(a).
The fiber has an elliptical germanium-doped core (doping level in the core equals \SI{18}{\mole\percent}) with a \SI{4.5}{\micro\meter} major axis and a \SI{2.9}{\micro\meter} minor axis.
The core is surrounded by rings of air holes arranged in honeycomb lattice. The lattice pitch of the air hole structure is \SI{3.0}{\micro\meter} on average and the diameter of the holes nearest to the core is \SI{0.9}{\micro\meter} on average.

We used cut-back technique to measure attenuation, Fig.~\ref{fig:fiber}(b). It does not exceed \SI{1}{dB\per\meter} at wavelengths shorter than \SI{2.115}{\micro\meter}.
Using white-light interferometric method, we measured relative inverse group velocities (RIGV, $\Delta\beta_1^{(x/y)}$) of the two polarization modes in a short fiber section, Fig.~\ref{fig:fiber}(c,d).
The results allowed to determine the value of group birefringence as approximately \SI{-2e-4}{} in the spectral region \SI{1.6}{}--\SI{2.15}{\micro\meter}.
From the RIGV slope we established chromatic dispersion. For pulses polarized along the slow fiber axis it is \SI{22}{\pico\second\per\kilo\meter\per\nano\meter} at the pump wavelength.

The fiber's geometrical parameters obtained from SEM image were used in the numerical modeling of the fiber transmission properties developed with Comsol Multiphysics Wave Optics Module.
Simulations performed using finite element method allowed to calculate the wavelength dependence of the effective refractive index and mode effective area $A_\mathrm{eff}$, Fig.~\ref{fig:fiber}(a).

To get an insight into the evolution of the pulse along the propagation distance we performed numerical simulations with a self-developed solver\footnote{https://github.com/WUST-FOG/cgnlse-python} based on the software implemented by Dudley~\cite{travers_frosz_dudley_2010}.
The simulations were based on the set of two-mode coupled nonlinear Schr\"odinger equations (CNLSE) in form
\begin{equation}
    \frac{\partial \tilde C_{x/y}}{\partial z} =
    D_{x/y}\left(\tilde C_{x/y}\right) + N_{x/y}\left(\tilde C_{x}, C_{y}\right),
    \label{eqn:CNLSE}
\end{equation}
where $\tilde C_{x/y}$ is a pulse pseudo-envelope written in the frequency domain for fast and slow polarization, respectively.
The dispersion operators for both polarizations were constructed by integrating measured RIGV $\Delta\beta_1^{(x/y)}\left(\omega\right)$, Fig.~\ref{fig:fiber}(c), and accounting for spectrally dependent loss $\alpha\left(\omega\right)$,
\begin{equation}
    D_{x/y}\left(\tilde C_{x/y}\right) =   \bigg(i\int_{\omega_0}^{\omega}\Delta\beta_1^{(x/y)}\left(\omega'\right)\mathrm{d}\omega' + \pm \Delta\beta_1^{(\mathrm{avg})}\left(\omega-\omega_0\right)
    -\frac{\alpha(\omega)}{2}\bigg) \tilde C_{x/y},
\end{equation}
where $\omega_0$ is the central pump frequency.
For the numerical simulations, we extrapolated loss dependencies to account short-wavelength Rayleigh scattering \cite{Saito2003, Sakaguchi:97} and water absorption~\cite{HUMBACH199619}.
The simulations were performed in the retarded time frame moving with the averaged group velocities of both polarization components at the pump frequency.

The spectral pseudo-envelope $\tilde C_{x/y}$ was introduced to account for the spectral dependency of effective mode area as proposed by Laegsgaard~\cite{Laegsgaard2007}.
As can be seen in Fig.~\ref{fig:fiber}(a), the effective area is almost identical regardless of the polarization, therefore we assumed the same value of $A_\mathrm{eff}(\omega)$ for both polarizations in the simulations.

Finally, the nonlinear terms in Eq.~(\ref{eqn:CNLSE}) take form:
\begin{multline}
    N_{x/y}\left(\tilde C_{x/y}, \tilde C_{y/x}\right) = \overline{\gamma_{x/y}}\mathcal{F}\Bigg\{
        \left(1-f_R\right)\times
        \left(\left|C_{x/y}\right|^2C_{x/y} + \frac{2}{3}\left|C_{y/x}\right|^2C_{x/y} + \frac{1}{3}C_{y/x}^2C_{x/y}^*\exp\left(\mp 2i\Delta\beta z\right)
        \right) + \\
        f_R\times\bigg[C_{x/y}\left(
            h_1\otimes\left|C_{x/y}\right|^2 +
            h_2\otimes\left|C_{y/x}\right|^2
            \right) + 
            C_{y/x}\left(h_3\otimes\bigg(
            C_{x/y}C_{y/x}^* +            C_{y/x}C_{x/y}^*\exp\left(\mp  2i\Delta\beta z\right)\right)
            \bigg)\bigg]\Bigg\}
\label{eqn:CNLSE_Nxy}    
\end{multline}
where $C_{x/y}$ denote temporal pseudo-envelopes of polarization modes, and $\overline{\gamma_{x/y}}$ are the effective nonlinear coefficients, $\Delta\beta$ is a difference between propagation constants of two polarization components at $\omega_0$, and $^*$ indicates conjugation.
The nonlinear terms account for the Raman and the Kerr nonlinearities, as described by Balla \textit{et al.}~\cite{Balla2018}. In Eq.~(\ref{eqn:CNLSE_Nxy}) \mbox{$f_R=0.245$} represents the fractional contribution of the delayed Raman response governed by the time-dependent functions $h_1(t)$, $h_2(t)$, and $h_3(t)$, definied as in~\cite{Martins2007,Rodrigo2013}.
For the propagation distance over \SI{1}{\meter}, we observed pulse overlap in the time window.
To avoid this issue, we filtered out soliton pulse in time domain after each meter of propagation. 

\begin{figure}[!b]
\centering
\includegraphics[width=0.8\linewidth]{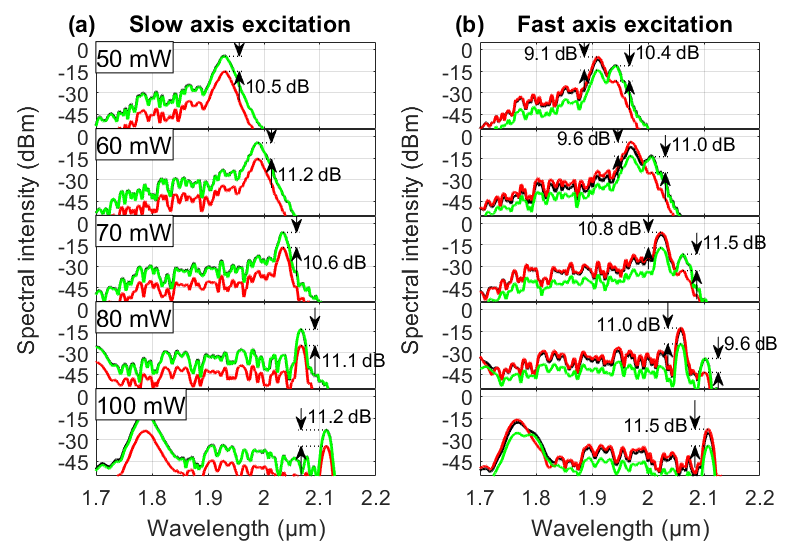}
\caption{\label{fig:spectra}Optical spectra recorded at the output of 20 meter long fiber section for increasing pump power. The polarization direction of the input pulse was aligned along the slow (a) and fast (b) axis of the fiber. We used polarizer to separate the signal components polarized along the slow (green) or fast (red) axis from the total (black) signal.
}
\end{figure}

Fig.~\ref{fig:spectra} shows the optical spectra recorded with an optical spectrum analyzer (Yokogawa AQ6376) at the output of a~20-meter long fiber section for different average power levels. The average power was measured before the lens used to couple light into the fiber.
The spectra were recorded with a bandpass filter (Thorlabs, FB2000-500) to reduce the intensity of the spectral region below \SI{1.75}{\micro\meter}.
We observe that the input pulse forms a~soliton, which undergoes Raman SSFS.
This process is schematically represented with an arrow in Fig.~\ref{fig:fiber}(c).
The characteristic frequency of maximal scalar Raman gain $\Delta\omega_{\parallel} \approx \SI{13}{THz}$~\cite{LinAgrawal2006}.
The soliton shift increases with the power but the soliton can be tuned hardly beyond \SI{2.1}{\micro\meter} due to the abrupt increase of attenuation, see Fig.~\ref{fig:fiber}(b).
When the input pulse is polarized along the slow axis, Fig.~\ref{fig:spectra}(a), the soliton pulse is generated with well-defined central wavelength.
However, when the input pulse is polarized along the fast axis, Fig.~\ref{fig:spectra}(b),
the main soliton pulse in fast axis is accompanied with orthogonally polarized components at longer wavelengths.
Due to the transmission loss of the fiber, we observed a decrease in solitons intensity above \SI{2.0}{\micro\meter}, moreover in case of \SI{100}{\milli\watt} average pump power, the trapped pulse is attenuated below noise level.
The polarization extinction ratio for all observed solitons ranges from \SI{9.1}{\dB} to \SI{11.5}{\dB}, Fig.~\ref{fig:spectra}.
There is no clear correlation between the polarization purity of the solitons generated and the input pulse power.
This suggests that polarization extinction is limited by linear coupling due to the imperfections of the fiber.

\begin{figure}[!h]
\centering
\includegraphics[width=0.8\linewidth]{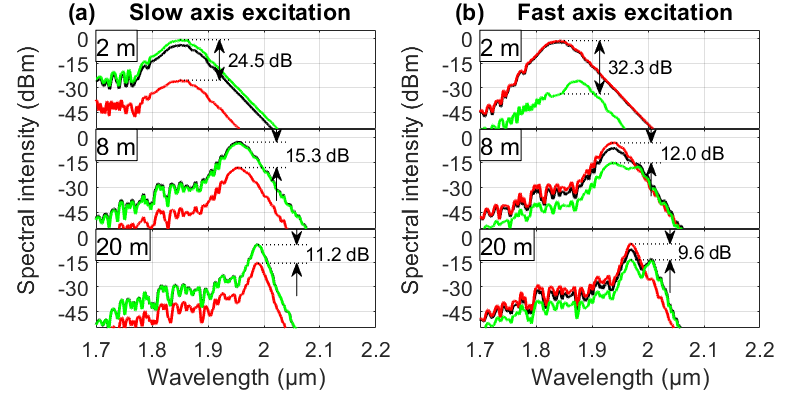}
\caption{Optical spectra recorded for pump power level of \SI{60}{mW} at the output of fiber sections of different lengths. The polarization direction of the input pulse was aligned along the slow (a) and fast (b) axis of the fiber. Line colors denote: green -- slow axis, red -- fast axis, black -- total spectrum.
\label{fig:spectra_length}
}
\end{figure}

In further study we focused on the spectral pulse dynamic with propagation distance, which was investigated via cutback of the original 20-m-long fiber.
The special care was taken in order to keep the fiber launch condition unchanged.
The spectra registered at the output of optical fiber sections of different lengths: \SI{2}{\meter}, \SI{8}{\meter}, and \SI{20}{\meter}; for the fixed average power of \SI{60}{\milli\watt} are presented in Fig.~\ref{fig:spectra_length}.
When the input pulse is polarized along the slow axis, there is no additional spectral component generated.
Contrary, when the input pulse is polarized along the fast axis, the slow axis component gradually increases.
This confirms that trapped pulse components in the slow axis appear due to orthogonal Raman scattering, which enables the simultaneous conversion of wavelength and polarization.
This process is schematically presented by the arrow in Fig.~\ref{fig:fiber}(d).
The characteristic frequency of maximal orthogonal Raman gain $\Delta\omega_{\perp} \approx \SI{3}{THz}$~\cite{LinAgrawal2006}.
In both excitation cases, the solitons shift toward longer wavelengths.
During propagation, they narrow and the extinction ratio lowers.
The spectral narrowing is direct consequence of the attenuation.
As the soliton energy decreases its duration increases and soliton narrows spectrally.

\begin{figure}[!h]
    \centering
    \includegraphics[width=0.61\linewidth]{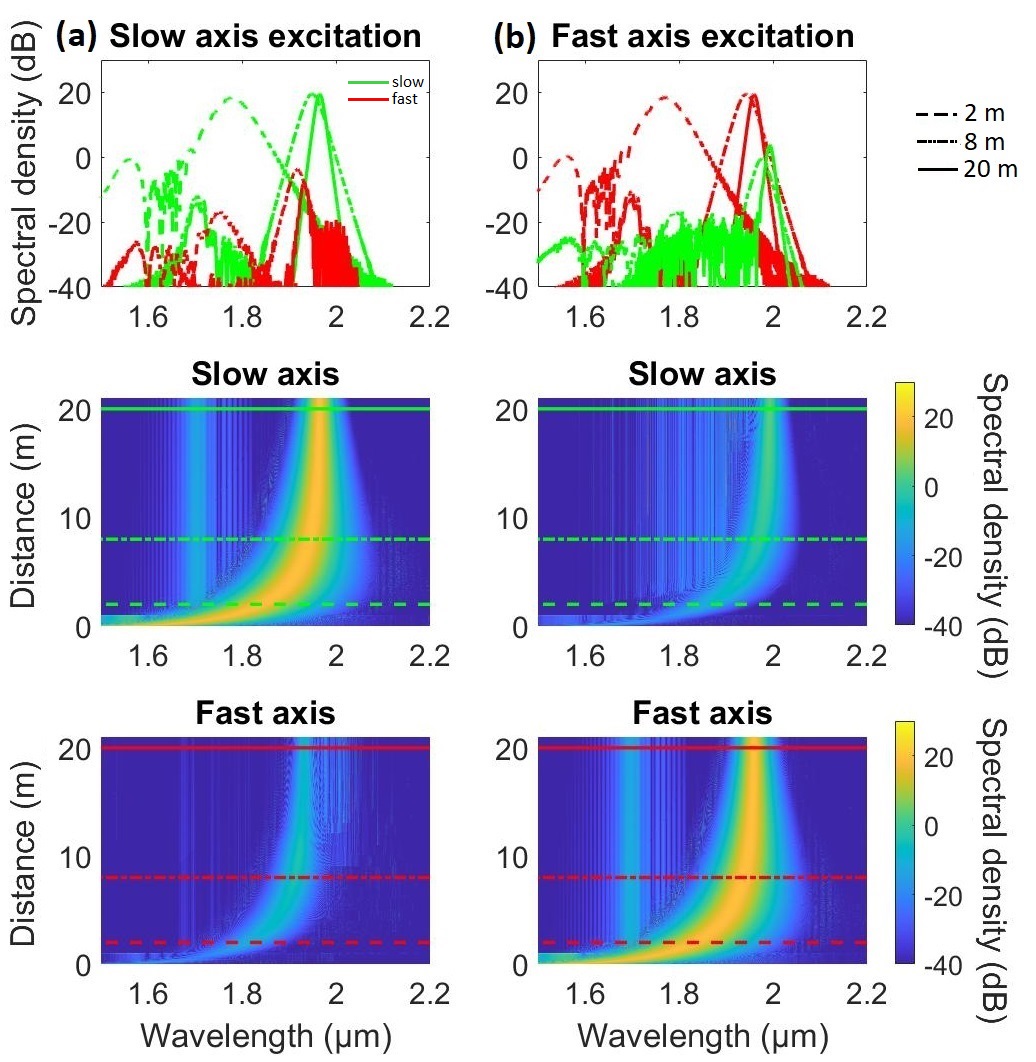}
    \caption{Calculated spectral evolution of generated soliton for polarization direction of input pulse aligned along the slow (a), and fast (b) axis of the fiber for \SI{60}{mW} pump power assuming 68\% coupling efficiency. Slow (fast) axis spectra are shown in green (red). Line styles correspond to propagation distances: \SI{2}{m}, \SI{8}{m}, \SI{20}{m}.}
    \label{fig:sim}
\end{figure}

Simulated spectral dynamics of generated pulse in both polarizations for two considered excitation conditions are presented in Fig.~\ref{fig:sim}.
The difference between the measurements and simulations is related to the linear coupling of polarizations which was not accounted for in the simulations.
Consequently, the polarization extinction ratio is higher than observed in experiments.
According to the simulation results, both polarization components are present in propagating pulse. 
In both excitation cases the fast axis component is located at shorter wavelength and the slow axis component is located at longer wavelength.
However, the orthogonal Raman scattering enables the transfer of energy only in one direction from shorter to longer wavelength.

\begin{figure}[!h]
\centering\includegraphics[width=0.63\linewidth]{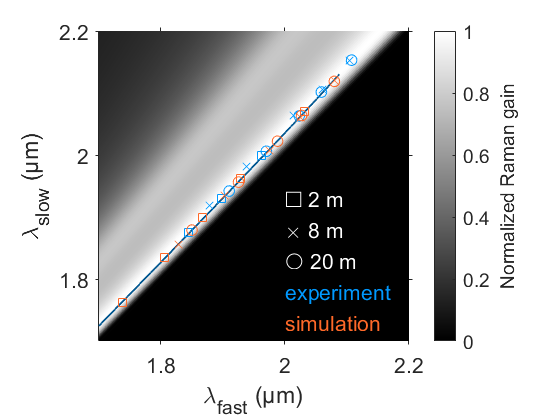}
\caption{\label{fig:GVM}Central wavelengths of two polarization components of the soliton pulse observed after different propagation distance when the input pulse is polarized along the fast axis (markers) and group velocity matched wavelengths for two fiber polarization modes (line). The background intensity indicates normalized orthogonal Raman gain from the fast-polarized pulse component (calculated after \cite{LinAgrawal2006}).}
\end{figure}

\newpage
Finally, we analyzed the relation between spectral positions of the main and trapped pulses for fast axis excitation.
We summarize the experimental and numerical results achieved for different propagation distances and power levels in Fig.~\ref{fig:GVM}.
Additionally, we plot the group velocity matched wavelengths (obtained from the RIGV of both polarization modes).
The markers indicating the central wavelengths of the pulse polarization components follow group velocity matching line, which indicates that the orthogonally polarized components have matched group velocities.
We note that the group velocity matching was also satisfied in the self-mode conversion investigated by Rishøj et al.~\cite{Rishoj2019}.
In the case of the  polarization conversion reported here, the process takes place on longer distance than in the mode conversion.
This is due to the smaller gain of orthogonal Raman scattering in comparison to the gain of co-polarized Raman scattering.
In background of Fig.~\ref{fig:GVM}, we plot the grayscale map of normalized orthogonal Raman gain calculated using Raman response function~\cite{LinAgrawal2006}.
It shows that in the fiber under consideration, the pulse components polarized along the slow axis are located at the spectral region for which the strongest orthogonal Raman gain is obtained from the soliton polarized along the fast axis.
This confirms that the slow-polarized component originates from the soliton formed initially in fast polarization.

In this work, we report observation of soliton trapping in microstructured birefringent fiber.
We confirmed that orthogonal Raman scattering provides the mechanism of energy transfer between polarization components.
Due to the negative sign of group birefringence the slow axis pumping assures higher polarization purity of tuned soliton than the fast axis pumping.
This result is important for the tunable sources of polarized light using Raman SSFS in the microstructured birefringent fibers~\cite{Sobon2018}.
In the next step, we plan to investigate the influence of group birefringence value on the soliton trapping.
Moreover, by controlling azimuth angle of linearly polarized light it is possible to control division of pulse power between polarization modes.
We presume that the polarization conversion efficiency strongly depends on the initial excitation of the fiber.

\textbf{Funding}
National Science Centre of Poland (DEC-2018/30/E/ST7/00862).

\bibliographystyle{unsrt}

\end{document}